\documentclass[aps,superscriptaddress,twocolumn,floatfix,preprintnumbers,nofootinbib]{revtex4}
\usepackage{amsmath,multirow,graphicx,tabularx}

\newcommand\one{\leavevmode\hbox{\small1\normalsize\kern-.33em1}}

\newcommand{\gev}{{\ensuremath\rm GeV}}

\newcommand{\ifb}{{\ensuremath\rm fb^{-1}}}

\def\slashchar#1{\setbox0=\hbox{$#1$}           
   \dimen0=\wd0                                 
   \setbox1=\hbox{/} \dimen1=\wd1               
   \ifdim\dimen0>\dimen1                        
      \rlap{\hbox to \dimen0{\hfil/\hfil}}      
      #1                                        
   \else                                        
      \rlap{\hbox to \dimen1{\hfil$#1$\hfil}}   
      /                                         
   \fi}

\def\eg{{\sl e.g.} \,}

\def\etal{{\sl et al} \,}

\newcommand{\be}{\begin{eqnarray*}}
\newcommand{\ee}{\end{eqnarray*}}

\newcommand{\bee}{\begin{eqnarray}}
\newcommand{\eee}{\end{eqnarray}}
\newcommand{\beeq}{\begin{equation}}
\newcommand{\eeeq}{\end{equation}}

\begin{document}

\title{Measuring Higgs Couplings at a Linear Collider}

\author{Markus Klute}
\affiliation{Massachusetts Institute of Technology, Cambridge, US}

\author{R\'emi Lafaye}
\affiliation{LAPP, Universit\'e Savoie, IN2P3/CNRS, Annecy, France}

\author{Tilman Plehn}
\affiliation{Institut f\"ur Theoretische Physik, Universit\"at Heidelberg, Germany}

\author{Michael Rauch}
\affiliation{Institute for Theoretical Physics, Karlsruhe Institute of Technology (KIT), Germany}

\author{Dirk Zerwas}
\affiliation{LAL, IN2P3/CNRS, Orsay, France}

\begin{abstract}
  Higgs couplings can be measured at a linear collider with high
  precision. We estimate the uncertainties of such measurements,
  including theoretical errors. Based on these results we show an
  extrapolation for a combined analysis at a linear collider and a
  high-luminosity LHC.
\end{abstract}

\maketitle

A new particle compatible with a Higgs boson~\cite{higgs} has been
discovered by ATLAS~\cite{atlas} and CMS~\cite{cms}. While the
existence of a narrow light resonance has been established beyond
reasonable doubt, the endeavor to study its properties has only
begun~\cite{cms_extra}.

Soon, the LHC energy will be increased to up to 14~TeV.  Detector and
machine upgrades can give us an integrated luminosity of up to $3000~\ifb$
at this energy, defining a high-luminosity LHC (HL-LHC) scenario.  A
linear collider (ILC) first accumulating $250~\ifb$ of data at 250~GeV
(LC250)~\cite{ILC,LEP3}, and upgradable to $500~\ifb$ at 500~GeV (LC500)~\cite{ILC}, can be
viewed as a dedicated Higgs factory. Further steps in center-of-mass energy
above 1~TeV are possible~\cite{Linssen:2012hp}.

We study the determination of the Higgs boson couplings at a linear
collider running at energies up to 500~GeV and combine its expected results
with extrapolated measurements at the HL-LHC.\bigskip

\underline{Analysis setup} --- Higgs couplings are defined as
prefactors of the respective Lagrangian terms coupling the Higgs field
to other Standard Model particles. They are defined relative to the
tree-level couplings predicted by the Standard
Model~\cite{sfitter_higgs,others},
\begin{alignat}{9}
g_{xxH} &\equiv g_x  = 
\left( 1 + \Delta_x \right) \;
g_x^\text{SM} \; .
\label{eq:delta}
\end{alignat}
The loop-induced Higgs-photon (Higgs-gluon) coupling then reads
\begin{alignat}{9}
g_{\gamma\gamma H} &\equiv g_{\gamma}  = 
\left( 1 + \Delta_\gamma^\text{SM} + \Delta_\gamma \right) \;
g_\gamma^\text{SM} \; .
\label{eq:deltagamma}
\end{alignat}
A modification of the underlying tree-level couplings induces
$\Delta_\gamma^\text{SM}$. The remaining $\Delta_\gamma$ characterizes
genuine non-Standard Model contributions. Equivalent parameters
$\kappa_x \equiv 1+\Delta_x$ have been introduced in
Ref.~\cite{HiggsXS}, without disentangling modified tree-level
couplings and new states in the loop-induced couplings.\medskip

All extrapolated HL-LHC measurements are based on the previous
detailed studies~\cite{sfitter_higgs,duehrssen}. They include a subjet
analysis for associated $VH$ production, but not the subjet analysis
in the $t\bar{t}H$ channel~\cite{boosted}.  The $VH$ channel is
crucial for the determination of
$\Delta_b$~\cite{sfitter_higgs,lecture}.

All statistical errors are scaled to the increased integrated luminosity.  The
statistical component of experimental uncertainties on background
rates, which are determined from data, will improve correspondingly.
The increase of statistics will also improve the statistical component
of the systematic errors. On the other hand, experimental conditions
(pile-up) will become significantly more difficult for some of the
crucial channels, like weak-boson-fusion Higgs production and hadronic
Higgs decays.
Therefore, the same performance of particle identification and
$b$-tagging as for lower instantaneous luminosity is assumed, i.e., the
relative errors for experimental systematics used in the previous
studies are not changed. 
Theory errors on the cross sections and on the Higgs branching ratios
are added linearly and included via the profile likelihood Rfit
scheme~\cite{rfit,sfitter}. Because Higgs analyses start to depend
more on exclusive jet observables~\cite{jetveto}, for which collinear factorization
may no longer hold, rendering the application of fixed-order QCD
corrections difficult, we refrain from postulating an improved theory
uncertainty.

All linear collider measurements used in this study are taken from
Refs.~\cite{lc,AguilarSaavedra:2001rg,Djouadi:2007ik}, with the exception of the additional
measurement of the $W$-fusion process with a $H \to b\bar{b}$ decay at
250~GeV~\cite{Deschprivate}. The expected error on the luminosity measurement
of 0.3\%~\cite{BozovicJelisavcic:2010hy} is added to each measurement
(100\% correlated), but has no effect on the end result.
As the ILC will likely first run at
250~GeV and then at 500~GeV, the latter results will include all
250~GeV measurements.  Also for the ILC each observable, with the
exception of the inclusive Higgs-strahlung cross section measurement,
is the product of cross section times branching ratio, so we apply the
same Rfit procedure as for the HL-LHC.

For the linear collider production cross sections we assume an error
of 0.5\% for $ZH$ and $\nu \bar{\nu} H$ production and of 1\% for
$t\bar{t}H$ production. As a consequence, the theoretical errors for
HL-LHC measurements are dominated by the cross sections, whereas at
the ILC the error on the branching ratio is the limiting factor.\bigskip

\underline{Higgs width} --- the main task of Higgs analyses is the
precise simultaneous determination of the individual
couplings~\cite{james}. Hadron collider experiments cannot directly
measure a narrow Higgs width, so they cannot simultaneously constrain
the couplings and new contributions to the total width. In addition,
the Higgs decay to charm quarks is challenging, but important, because
it contributes to the total width at the per-cent level. Therefore, we
assume
\begin{alignat}{9}
\Gamma_\text{tot} = \sum_\text{obs} \; \Gamma_x(g_x) 
+ \text{2nd generation} < 2\,\gev \; .
\label{eq:width}
\end{alignat}
The upper limit of 2~GeV takes into account that a larger width would
become visible in the mass measurement. The second generation is
linked to the third generation via $g_c = m_c/m_t \; g_t^\text{SM} (1
+ \Delta_t)$. The leptonic muon Yukawa might be observable at the LHC
in weak boson fusion or inclusive searches, depending on the available
luminosity~\cite{muons}.\medskip

\begin{figure}[t]
\includegraphics[width=0.90\columnwidth]{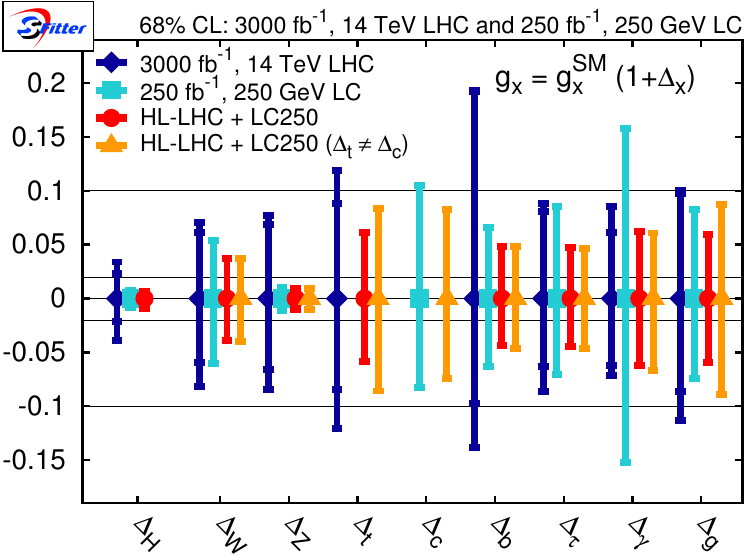}
\caption{\label{fig:LC250} Expected precision for Higgs coupling
  measurements at the HL-LHC, ILC at 250~GeV and their
  combination. For the latter we also show the fit including
  $\Delta_c$. The inner bars for HL-LHC denote a scenario with improved
  experimental systematic uncertainties.}
\end{figure}

\begin{figure}[t]
\includegraphics[width=0.90\columnwidth]{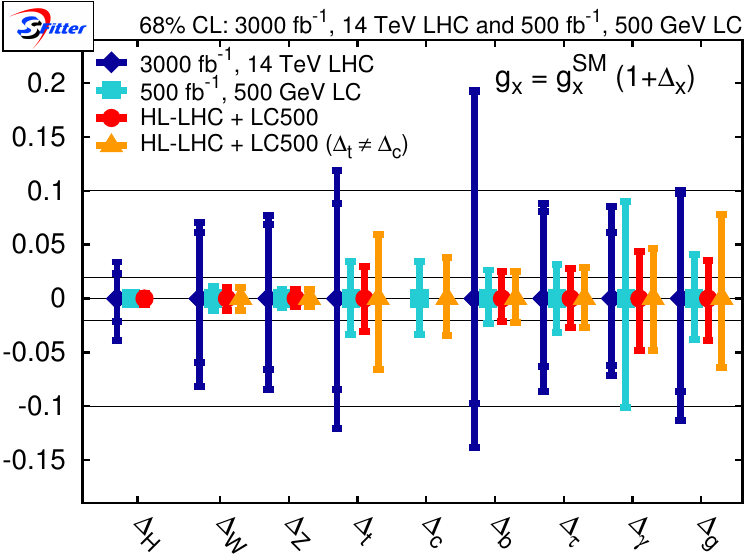}
\caption{\label{fig:LC500} Expected precision for Higgs couplings
  measurements at the HL-LHC, ILC up to 500~GeV and their
  combination. For the latter we also show the fit including
  $\Delta_c$. The inner bars for HL-LHC denote a scenario with improved
  experimental systematic uncertainties.}
\end{figure}
\begin{table*}[t]
\begin{tabular}{l||c|c||c|c|c||c|c|c}
\hline
         & 
HL-LHC &
HL-LHC &
LC250 &
LHC + LC250 &
LHC + LC250 &
ILC500 &
LHC + LC500 &
LHC + LC500 \\
&
&
improved&
&
&
$\Delta_c \ne \Delta_t$ &
&
&
$\Delta_c \ne \Delta_t$ \\\hline
\multirow{2}{*}{$\Delta_H$}
& $-0.04$ & $-0.02$ & $-0.009$ & \multicolumn{2}{c||}{$-0.009$} & $-0.005$ & \multicolumn{2}{c}{$-0.006$} \\
& $+0.03$ & $+0.02$ & $+0.008$ & \multicolumn{2}{c||}{$+0.007$} & $+0.005$ & \multicolumn{2}{c}{$+0.005$} \\\hline
\multirow{2}{*}{$\Delta_W$}
& $-0.08$ & $-0.06$ & $-0.06~$ & $-0.04~$ & $-0.04~$ & $-0.011$ & $-0.011$ & $-0.011$ \\ 
& $+0.07$ & $+0.06$ & $+0.05~$ & $+0.04~$ & $+0.04~$ & $+0.011$ & $+0.010$ & $+0.010$ \\\hline
\multirow{2}{*}{$\Delta_Z$}
& $-0.08$ & $-0.07$ & $-0.011$ & $-0.010$ & $-0.010$ & $-0.008$ & $-0.008$ & $-0.008$ \\ 
& $+0.08$ & $+0.07$ & $+0.010$ & $+0.009$ & $+0.009$ & $+0.008$ & $+0.008$ & $+0.008$ \\\hline
\multirow{2}{*}{$\Delta_t$}
& $-0.12$ & $-0.08$ & \multirow{2}{*}{ --- } &          & $-0.09~$ &          &          & $-0.07~$ \\ 
& $+0.12$ & $+0.09$ &                        & $-0.06~$ & $+0.08~$ & $-0.03~$ & $-0.03~$ & $+0.06~$ \\\cline{1-4}\cline{6-6}\cline{9-9}
\multirow{2}{*}{$\Delta_c$}
& \multirow{2}{*}{ --- } & \multirow{2}{*}{ --- } & $-0.08~$ & $+0.06~$ & $-0.07~$ & $+0.04~$ & $+0.03~$ & $-0.03~$ \\ 
&                        &                        & $+0.11~$ &          & $+0.08~$ &          &          & $+0.04~$ \\\hline
\multirow{2}{*}{$\Delta_b$}
& $-0.14$ & $-0.10$ & $-0.06~$ & $-0.04~$ & $-0.05~$ & $-0.02~$ & $-0.02~$ & $-0.02~$ \\ 
& $+0.20$ & $+0.19$ & $+0.07~$ & $+0.05~$ & $+0.05~$ & $+0.03~$ & $+0.03~$ & $+0.03~$ \\\hline
\multirow{2}{*}{$\Delta_\tau$}
& $-0.09$ & $-0.06$ & $-0.07~$ & $-0.05~$ & $-0.05~$ & $-0.03~$ & $-0.03~$ & $-0.03~$ \\ 
& $+0.09$ & $+0.08$ & $+0.09~$ & $+0.05~$ & $+0.05~$ & $+0.03~$ & $+0.03~$ & $+0.03~$ \\\hline
\multirow{2}{*}{$\Delta_\gamma$}
& $-0.07$ & $-0.06$ & $-0.15~$ & $-0.06~$ & $-0.07~$ & $-0.10~$ & $-0.05~$ & $-0.05~$ \\ 
& $+0.09$ & $+0.06$ & $+0.16~$ & $+0.06~$ & $+0.06~$ & $+0.09~$ & $+0.04~$ & $+0.05~$ \\\hline
\multirow{2}{*}{$\Delta_g$}
& $-0.11$ & $-0.09$ & $-0.08~$ & $-0.06~$ & $-0.09~$ & $-0.04~$ & $-0.04~$ & $-0.06~$ \\ 
& $+0.10$ & $+0.10$ & $+0.08~$ & $+0.06~$ & $+0.09~$ & $+0.04~$ & $+0.04~$ & $+0.08~$ 
\end{tabular}
\caption{Errors on Higgs couplings for different collider scenarios,
  as shown in Figs.~\ref{fig:LC250} and \ref{fig:LC500}.}
\label{tab:errors}
\end{table*}

At the ILC the situation is very different: the total width can be
inferred from a combination of measurements. This is mainly due to  
the measurement of the inclusive $ZH$ cross section based
on a system recoiling against a $Z \to \mu^+\mu^-$ decay. 
While the simultaneous fit of all couplings will reflect
this property, we can illustrate this feature based on four
measurements~\cite{AguilarSaavedra:2001rg,Djouadi:2007ik}
\begin{enumerate}
\item Higgs-strahlung inclusive ($\sigma_{ZH}$)
\item Higgs-strahlung with a decay to $b\bar{b}$ ($\sigma_{Zbb}$)
\item Higgs-strahlung with a decay to $WW$ ($\sigma_{ZWW}$)
\item $W$-fusion with a decay $b\bar{b}$ ($\sigma_{\nu\nu bb}$)
\end{enumerate}
described by four unknowns $\Delta_W$,
$\Delta_Z$, $\Delta_b$, and $\Gamma_\text{tot}$. Schematically, the
total width is
\begin{alignat}{5}
\Gamma_\text{tot} \leftarrow 
\dfrac{\sigma_{\nu\nu bb}/\sigma_{Zbb}}
      {\sigma_{ZWW}/\sigma_{ZH}}
\times \sigma_{ZH} \; .
\end{alignat}
This results in a precision of about 10\%~\cite{Deschprivate} on the total width
at LC250.

In addition, Higgs decays to charm quarks can be disentangled from the
background, therefore a link between the second and third generation
along the lines of Eq.\eqref{eq:width} is not needed. A difference in
the interpretation of our results we need to keep in mind: while
electroweak corrections are not expected to interfere at the level of
precision of our HL-LHC analysis, at the ILC the individual
measurement of Higgs couplings will most likely require an appropriate
ultraviolet completion~\cite{passarino}. In this largely
experimentally driven study we assume the existence of such a picture.

At a linear collider the errors on Higgs branching ratios $\text{BR}_x$
or particle widths $\Gamma_x$ are crucial~\cite{hdecay}. As theory
errors on the latter we assume 4\% for decays into quarks, 2\% for gluons, and 1\% for all other
decays~\cite{sfitter_higgs}. 
Translated into branching ratios this
corresponds for example to an error around 2\% on the branching ratio into bottom
quarks. Further improvements on these values in the
future are possible, but we decided to remain conservative.
The error on the branching ratios follows
from simple error propagation, where theory errors are added linearly,
\begin{alignat}{5}
\delta \text{BR}_{x} &= \sum_{k} \left| \frac{\partial}{\partial \Gamma_k}
\text{BR}_{x} \right| \delta \Gamma_k \notag \\
&= \frac1{\Gamma_\text{tot}} \left( \text{BR}_x \sum_{k} \delta \Gamma_k 
   + \left( 1 - 2 \text{BR}_x \right) \delta\Gamma_x \right) \; .
\end{alignat}
\bigskip

\underline{Higgs couplings} --- the result of an individual and
simultaneous determination of the Higgs couplings are shown in
Fig.~\ref{fig:LC250}. For the LHC, we need to make an assumption about
the width, shown in Eq.~\eqref{eq:width}. At LC250 the inclusive $ZH$
rate gives direct access to $\Delta_Z$ at the percent level. No assumption about the width
is needed. 

The simplest model for modified Higgs couplings is a global factor
$\Delta_H$, which arises through a Higgs portal~\cite{portal} or in
simple strongly interacting extensions~\cite{silh}. In
Fig.~\ref{fig:LC250} we see that we can measure this single parameter
at the HL-LHC with an error around 4\%. A further increase in
statistics would not improve this error as this determination is
limited by the theoretical error. Reducing the luminosity error from 5\%
to 2\% and neglecting all other systematic errors, the accuracy 
improves only by about 20\% as shown by the inner bars, corroborating
the theory limitation.
The LC250 with its smaller theoretical error improves the determination
to about 1\% with a further decrease in the error for the LC500 as shown in
Fig.~\ref{fig:LC500}.  The combination of measurements from HL-LHC with
LC250 or LC500 is dominated by the precision of the linear collider.  All numerical
values in Figs.~\ref{fig:LC250} and~\ref{fig:LC500} are also reproduced
in Table~\ref{tab:errors}.\medskip
 
The (more) model independent determination of individual Higgs
couplings is also shown in Fig.~\ref{fig:LC250}.  At the HL-LHC alone
the most precisely measured couplings are those to the weak bosons at
about 8\%. The measurement of the quark Yukawas is challenging, as
reflected in the larger errors. Unless one of the ingredients to this
analysis significantly changes, LHC will measure Higgs couplings to
Standard Model particles at an excellent $8-15\%$ level, assuming no
significant contributions from new physics channels~\footnote{For this
  specific LHC scenario we estimate the effect of flat theory
  errors~\cite{rfit,sfitter} compared to Gaussian theory errors. For
  the same width we find that a Gaussian ansatz underestimates the
  errors bars on the $\Delta_j$ by a relative 10\% to 20\%.}.

For easier comparison in Fig.~\ref{fig:LC250} and Fig.~\ref{fig:LC500} 
the constraint on the Higgs width and the link between the second and third generation was applied 
also for LC250 and LC500.
However, once linear collider data enter the picture, these two 
assumptions can be removed. 
This introduces $\Delta_c$ is as an additional parameter. The impact of the 
additional parameter on the results is illustrated only in the combined analyses.

Compared to the individual results for the HL-LHC or the LC250, a significant
improvement of the joint analysis is clearly visible. For example,
$\Delta_W$ is improved in the combined analysis; according to
Eq.\eqref{eq:deltagamma} this improves the measurement of
$\Delta_\gamma^\text{SM}$, so $\Delta_\gamma$ can be determined at the
5\% level.
When adding $\Delta_c$, the precision decreases slightly, shown in Fig.~\ref{fig:LC250}, as expected 
from adding an additional parameter to a system while using the same observables.
The limitation of the LC250 coupling determination comes from the
total width, which can be measured to 10\%~\cite{Deschprivate} due to the small cross section
for $W$-fusion at this center-of-mass energy. As a matter of fact, at the LC250 all
couplings with the exception of $\Delta_b$ are limited by
statistics.\medskip

Once the ILC energy is increased to 500~GeV, more decay channels can
be observed in $W$-fusion production.  The results of the coupling
measurement are shown in Fig.~\ref{fig:LC500}. Compared to the LC250
setup the precision on the tree-level couplings is improved by roughly
a factor~2 and $\Delta_t$ is now measured directly. The overall
picture of the combined HL-LHC and LC500 analysis does not change
qualitatively: in particular the loop-induced couplings benefit
significantly when compared to individual linear collider and HL-LHC
analyses.\bigskip

\underline{Conclusions} --- The measurement of Higgs couplings is a
prime objective for a linear collider. While a HL-LHC can reach a precision of
the order of 8\%, the linear collider measurements will improve on this precision
by almost an order of magnitude.  Even more importantly, linear collider
measurements provide direct access the total Higgs width and to
numerically relevant second-generation couplings like $\Delta_c$. The
study of exotic Higgs decays is not any longer limited by the ability
to identify the corresponding final state.

A combined analysis of HL-LHC and ILC measurements will make the
determination of the couplings more precise than at any single
machine.  HL-LHC and ILC form a dream team to study the properties of
the Higgs boson and to establish the properties of the Higgs boson in
and beyond the Standard Model with high precision.\medskip

\underline{Acknowledgments} --- We would like to thank 
Klaus Desch, Keisuke Fujii, Michael Peskin and Peter Zerwas for helpful discussions.
MR acknowledges support by the Deutsche Forschungsgemeinschaft via the
Sonderforschungsbereich/Transregio SFB/TR-9 ``Computational Particle
Physics'' and the Initiative and Networking Fund of the Helmholtz
Association, contract HA-101(``Physics at the Terascale'').  Part of
the work was performed in the GDR Terascale of the CNRS/IN2P3.


\end{document}